# Consistent Streaming Through Time: A Vision for Event Stream Processing

Roger S. Barga, Jonathan Goldstein, Mohamed Ali and Mingsheng Hong.

Microsoft Research
Redmond, WA
*{barga, jongold,t-mohali,t-minhon} @microsoft.com*

## ABSTRACT
Event processing will play an increasingly important role in constructing enterprise applications that can immediately react to business critical events. Various technologies have been proposed in recent years, such as event processing, data streams and asynchronous messaging (e.g. pub/sub). We believe these technologies share a common processing model and differ only in target workload, including query language features and consistency requirements. We argue that integrating these technologies is the next step in a natural progression. In this paper, we present an overview and discuss the foundations of CEDR, an event streaming system that embraces a temporal stream model to unify and further enrich query language features, handle imperfections in event delivery, define correctness guarantees, and define operator semantics. We describe specific contributions made so far and outline next steps in developing the CEDR system.

## Categories and Subject Descriptors
H.1.1 [Systems and Information Theory]: General Systems Theory

## 1. Motivation and Introduction

Most businesses today actively monitor data streams and application messages, in order to detect *business events* or *situations* and take time-critical actions [1]. It is not an exaggeration to say that business events are the real drivers of the enterprise today because they represent changes in the state of the business. Unfortunately, as in the case of data management in pre-database days, every usage area of business events today tends to build its own special purpose infrastructure to filter, process, and propagate events.

Designing efficient, scalable infrastructure for monitoring and processing events has been a major research interest in recent years. Various technologies have been proposed, including data stream management, complex event processing, and asynchronous messaging such as pub/sub. We observe that these systems share a common processing model, but differ in query language features. Furthermore, applications may have different requirements for *consistency*, which specifies the desired tradeoff between insensitivity to event arrival order and system performance. Clearly, some applications require a strict notion of correctness that is robust relative to event arrival order, while others are more concerned with high throughput. If exposed to the user and handled within the system, users can specify consistency requirements on a per query basis and the system can adjust consistency at runtime to uphold the guarantee and manage system resources.

To illustrate, consider a financial services organization that actively monitors financial markets, individual trader activity and customer accounts. An application running on a trader's desktop may track a moving average of the value of an investment portfolio. This moving average needs to be updated continuously as stock updates arrive and trades are confirmed, but does not require perfect accuracy. A second application running on the trading floor extracts events from live news feeds and correlates these events with market indicators to infer market sentiment, impacting automated stock trading programs. This query looks for patterns of events, correlated across time and data values, where each event has a short "shelf life". In order to be actionable, the query must identify a trading opportunity as soon as possible with the information available at that time; late events may result in a retraction. While a third application running in the compliance office monitors trader activity and customer accounts, to watch for churn and ensure conformity with SEC rules and institution guidelines. These queries may run until the end of a trading session, perhaps longer, and must process all events in proper order to make an accurate assessment. These applications carry out similar computations but differ significantly in their workload and requirements for consistency guarantees and response time.

This example illustrates that most real-world enterprise applications are complex in functionality, and incorporate different technologies that must work together with strict requirements in terms of accuracy and consistency. We believe these technologies complement each other and will naturally converge in future systems, but several research and engineering challenges must first be addressed. We present our analysis on existing technologies as follows.

Data stream systems, which support sliding window operations and use sampling or approximation to cope with unbounded streams, could be used to compute a moving average of portfolio values. However, there are important features that cannot be naturally supported in existing stream systems. First, *instance selection and consumption* can be used to customize output and increase system efficiency, where selection specifies which event instances will be involved in producing output, and consumption specifies which instances will never be involved in producing future output, and therefore can be effectively "consumed". Without this feature, an operator such as sequence [13] is likely to be too expensive to implement in a stream setting – no past input can be forgotten due to its potential relevance to future output, and the size of output stream can be multiplicative w.r.t. the size of the input. Expressing negation or the non-occurrence of events, such as a customer not answering an email within a specified time, in a query is useful for many applications, but can not be naturally expressed in many existing stream systems. Messaging systems such as pub/sub, could handily route news feeds and market data but pub/sub queries are usually stateless and lack the ability to carry out computation other than filtering. Complex event processing systems can detect patterns in event streams, including both the occurrence and non-occurrence of events, and queries can specify intricate temporal constraints. However, most event systems available today provide only limited support for *value constraints or correlation* (predicates on event attribute values), as well as query directed instance selection and consumption policies. Finally, none of the above technologies provide support for consistency guarantees.

We contend that data streams, complex event processing and pub/sub are complementary technologies and propose a paradigm that integrates and extends these models, and upholds precise notions of consistency. We are developing a system called CEDR (*Complex Event Detection and Response*) to explore the benefits of an event streaming system that integrates the above technologies, and supports a spectrum of consistency guarantees. This paper presents a current snapshot of the CEDR project. We are not presenting a complete system at this time as several research and engineering challenges remain. However, there are a number of concrete contributions to report on at this point:

- A stream data model that embraces a temporal data perspective, and introduces a clear separation of different notions of time in streaming applications (Section 2).

- A declarative query language capable of expressing a wide range of event patterns with temporal and value correlation, negation, along with query directed instance selection and consumption. All aspects of the language are fully composable (Section 3).

- Along with the language, we define a set of logical operators that implement the query language, and serve as the basis for logical plan exploration during query optimization.

- We formally define a spectrum of consistency levels to deal with stream imperfections, such as latency or out-of-order delivery, and to meet application requirements for quality of the result. We also discuss the consequences of upholding the consistency guarantees in a streaming system (Sections 4 and 5).

- We base our implementation on a set of run-time operators, most of which are based on view update semantics. We provide the denotational semantics of these operators, and formally define notions of good behavior and view update compliance. We also introduce a novel operator, called AlterLifetime, which can be used to implement a variety of window types (Section 6).

Due to space limitations, we do not include a section dedicated to related work, but refer the reader to our technical report [2] which includes a discussion of related work. We do make comparisons to systems throughout this paper, particularly *STREAM* [5], *Aurora* [4], Niagra [9] Nile [10], Cayuga [7] and HiFi [3]. However even these comparisons are narrowly focused and again we refer the reader to [2].

## 2. CEDR Temporal Stream Model

In this section, we introduce our tritemporal stream model, the theoretical foundation for CEDR which allows us to support both query language semantics and consistency guarantees simultaneously. Existing stream systems already separate the notion of application time and system time [11], where the former is the clock that event providers use to timestamp tuples they generate, and the latter is the clock of the stream processing server. In CEDR, we further refine application time into two temporal dimensions, valid time and occurrence time, and refer to system time as CEDR time. This gives us three temporal dimensions in our stream model. We now describe each notion of time in detail.

In CEDR, a data stream is modeled as a time varying relation. Each tuple in the relation is an event, and has an ID. Each tuple has a *validity interval*, which indicates the range of time when the tuple is valid from the event provider's perspective. Given the interval representation of each event, it is possible to issue the following continuous query: "at each time instance t, return all tuples that are still valid at t." Note that existing systems [4, 5] model stream tuples as points, and therefore do not capture the notion of validity interval. Consequently, they cannot naturally express such a query. An interval can be encoded with a pair of points, but the resulting query formulation will be unintuitive.

After an event initially appears in the stream, we allow its validity interval (e.g. the time during which a coupon could be used) to be changed by the event provider, a feature not naturally supported in existing stream systems. Such changes are represented by tuples with the same ID but different content. A second temporal dimension, *occurrence time*, models when such changes occur from the event provider's perspective. An *insert* event of a certain ID is the tuple with minimum occurrence start time value ($O_s$) among all events with that ID. Other events of the same ID are referred to as *modification* events. Both valid time and occurrence time are assigned by the same logical clock of the event provider, and are thus comparable[1]. We use $t_v$ to denote valid time, and use $t_o$ to denote occurrence time.

We use the following schema as the conceptual representation of a stream produced by an event provider: (ID, $V_s$, $V_e$, $O_s$, $O_e$, Payload). Here $V_s$ and $V_e$ respectively denote valid start and end time; $O_s$ and $O_e$ respectively denote occurrence start and end time; Payload is the sub-schema consisting of normal value attributes, and is application dependent. For example, Figure 1 represents the following scenario: at time 1, event e0 is inserted into the stream with validity interval [1, ∞); at time 2, e0's validity interval is modified to [1, 10]; at time 3, e0's validity interval is modified to [1, 5), and e1 is inserted with validity interval [4, 9). We ignore the content payload in examples throughout this paper, and focus only on temporal attributes.

**Figure 1. Example – Conceptual stream representation**

| ID | $V_s$ | $V_e$ | $O_s$ | $O_e$ | (Payload) |
|----|-------|-------|-------|-------|-----------|
| e0 | 1 | ∞ | 1 | 2 | … |
| e0 | 1 | 10 | 2 | 3 | … |
| e0 | 1 | 5 | 3 | ∞ | … |
| e1 | 4 | 9 | 3 | ∞ | … |

We stress that the bitemporal schema above is only a conceptual representation of a stream. In an actual implementation, stream schemas can be customized to fit application scenarios. This is similar to the notion of temporal specialization in the literature [12]. When events produced by the event provider are delivered into CEDR, they can become out of order, due to unreliable network protocols, system crash recovery, and other anomalies in the physical world. We model out-of-order event delivery with a third temporal dimension, producing a tritemporal stream model. This is further discussed in Section 4.

# 3. CEDR Query Language

CEDR query semantics are defined only on the information obtained from event providers, and this implies the query language will reason about valid and occurrence time, but not CEDR time. When we specify the semantics of a CEDR query, its input and output are both bitemporal streams (consisting of valid time and occurrence time).

The CEDR language for registering event queries is based on the following three aspects: 1) *event pattern expression*, composed by a set of high level operators that specify how individual events are *filtered*, and how multiple events are *correlated (joined)* via time-based and value-based constraints to form *composite event instances*, or *instances* for short. 2) *Instance selection and consumption*, expressed by a policy referred to as an SC mode; 3) finally, *instance transformation*, which takes the events participating in a detected pattern as input, and transforms them to produce complex output events via mechanisms such as aggregation, attribute projection, and computation of a new function. In designing the CEDR language, we took efforts to make sure that all features are fully composable with each other.

## 3.1 Overview of the CEDR Language

Due to space constraints, here we give an overview of the language syntax and semantics through a query example.

**EVENT** CIDR07_Example
**WHEN** UNLESS(SEQUENCE(INSTALL *x*,
            SHUTDOWN AS *y*, 12 hours),
        RESTART AS *z*, 5 minutes)
**WHERE** {*x.Machine_Id* = *y.Machine_Id*} AND
        {*x.Machine_Id* = *z.Machine_Id*}

The SEQUENCE construct specifies a sequence of events that must occur in a particular order. The parameters of the SEQUENCE operator (or any operator that produces composite events in general) are the occurrences of events of interest, referred to as *contributors*. There is a *scope* associated with the sequence operator, which puts an upper bound on the temporal distance between the occurrence of the last contributor in the sequence and that of the first contributor. In this query, the SEQUENCE construct specifies a sequence that consists of the occurrence of an *INSTALL* event followed by a *SHUTDOWN* event, within 12 hours of the occurrence of the former. The output of the SEQUENCE construct should then be followed by the non-occurrence of a *RESTART* event within 5 minutes. Non-occurrences of events, also referred to as *negation* in this work, can be expressed either directly using the NOT operator, or indirectly using the UNLESS operator, which is used in this query formulation. Intuitively, UNLESS(A, B, w) produces an output when the occurrence of an A event is followed by non-occurrence of any B event in the following w time units. w is therefore the negation scope. In this query, UNLESS is used to express that the sequence of INSTALL, SHUTDOWN events should not be followed by no RESTART event in the next 5 minutes. We can also bind a sub-expression to a variable via AS construct, such that we can refer to the corresponding contributor in WHERE clause when we specify value constraints.

---
[1] Valid and occurrence time can be assigned by different physical clocks, in which case we require them to be synchronized.

Now we continue to describe the WHERE clause for this query. There we use the variables defined previously to form predicates that compare attributes of different events. To distinguish from simple predicates that compare to a constant like those in the first example, we refer to such predicates as *parameterized predicates* as the attribute of the later event addressed in the predicate is compared to a value that an earlier event provides. The parameterized predicates in this query compare the *id* attributes of all three events in the WHEN clause for equality. Equality comparisons on a common attribute across multiple contributors are typical in monitoring applications. For ease of exposition, we refer to the common attribute used for this purpose as a correlation key, and the set of equality comparisons on this attribute as an *equivalence test*. Our language offers a shorthand notation: an equivalence test on an attribute (e.g., *Machine_Id*) can be expressed by enclosing the attribute name as an argument to the function CorrelationKey with a keywords, such as EQUAL, UNIQUE (e.g., CorrelationKey(*Machine_ID, Equal)*, as shown in the comment on the WHERE clause in this example). Moreover, if an equivalence test requires all events to have a specific value (e.g., 'BARGA_XP03') for the attribute *id*, we can express it as [*Machine_Id* Equal 'BARGA_XP03'].

Instance selection and consumption should be specified in WHEN clause as well. For simplicity of the query illustration, we did not show their corresponding syntax constructs in the above query, and will defer the description of SC modes supported in CEDR till a later point. Finally, instance transformation is specified in an optional OUTPUT clause to produce output events. If OUTPUT clause is not specified in a query, all instances that pass the instance selection process will be output directly to the user.

### 3.2 Features of CEDR Language

Due to space constraints, in this section we only highlight features that distinguish CEDR from other event processing and data stream languages.

**Event Sequencing** – The ability to synthesize events based upon the ordering of previous events is a basic and powerful event language construct. For efficient implementation in a stream setting, all operators that produce outputs involving more than one input event should have a time based scope, denoted as w. For example, SEQUENCE(E1, E2, w) outputs a sequence event at the occurrence of an E2 event, if there has been an E1 event occurrence in the last w time units. Most event processing systems, such as SNOOP [6], do not support scope. In Cayuga [7] and SASE [13], scope is expressed respectively by a duration predicate and a window clause. In CEDR, scope is "tightly coupled" with operator definition, and thus helps users in writing properly scoped queries, and permits the optimizer to generate efficient plans.

**Negation** – Negation has to have a scope within which the non-occurrence of events is monitored. The scope can be time based or sequence based. The CEDR language has three negation operators. We informally describe their semantics below. First, for time scope, UNLESS(E1, E2, w) produces an output event when the occurrence of an E1 event is followed by no E2 event in the next w time units. The start time of negation scope is therefore bound always to the occurrence of the E1 event. For sequence scope, we use the operator NOT (E, SEQUENCE (E1,…,Ek, w)), where the second parameter of NOT, a sequence operator, is the scope for the non-occurrence of E. It produces an output at the occurrence of the sequence event specified by the sequence operator, if there is no occurrence of E between the occurrence of E1 and Ek that contribute to the sequence event. Finally, CANCEL-WHEN (E1, E2) stops the (partial) detection for E1 when an E2 event occurs. Cancel-when is a powerful language feature not found in existing event or stream systems. Unlike existing systems [13], negation in CEDR is fully composable with other operators.

**Temporal Slicing** – We have two temporal slicing operators @ and # respectively on occurrence time and valid time. Users can put them in the query formulation to customize the bitemporal query output. For example, for Q @ [$t_{o1}$, $t_{o2}$] #[$t_{v1}$, $t_{v2}$]), among the tuples in the bitemporal output of query Q, it only outputs tuples valid between $t_{v1}$ and $t_{v2}$, and occur at time between $t_{o1}$ and $t_{o2}$.

**Value Correlation in the WHERE clause** – Some existing event languages [13] support WHERE clause. However, when the language supports negation, for a query in which negation is composed with other operators in a complex way, it could become quite hard to reason about the semantics of value correlation. In CEDR, we carefully define the semantics of such value correlation based on what operators are present in the WHEN clause, by placing the predicates from the WHERE clause into the denotation of the query, a process we refer to as *predicate injection*. SASE [13] takes a simpler approach, since the language operators in SASE are not composable. Overall, predicate injection for negation is non-trivial, and is simply not handled by many existing systems.

**Instance Selection and Consumption** – Many systems do not support this feature [13], while others tailor the semantics of instance selection and consumption in favor of theoretical properties, and are thus "arbitrary" from a user's perspective; i.e., not controlled by user on a per query basis. In some cases, the semantics of selection and consumption are "hard coded" into operator definitions, and thus inflexible [7, 8]. In CEDR the specification of SC mode is decoupled from operator semantics, and for language composability, SC mode is associated with the input parameters of operators, instead of only base stream events.

## 3.3 Formal Language Semantics

In order for a query language to be compositional, the type of the query output should be the same as that of the query inputs. Hence, in the case of bitemporal databases and CEDR streams, the output type of a query should be a bitemporal relation. We now formally define the semantics of the CEDR language constructs with the *denotation* in relational calculus style. First, we focus on operators used in the WHEN clause. In many event processing systems, low level event algebra operators are the *only way* to specify a complex event pattern for detection. The functionality or meaning of these operators is not always intuitive, leading to confusion and documented peculiarities and irregularities. Our approach is to provide high level operators with intuitive and well-defined semantics. Operators can be composed to form an *event expression* in the WHEN clause. To make the operators composable, each input parameter of an operator is itself an event expression. The simplest event expression is an event type, which outputs all events of this event type. Below, we describe the set of operators that CEDR supports and formally present their semantics.

### 3.3.1 Conventions

Each event is associated with a type, and has a header and a body component in its content. The header consists of temporal attributes, the ID column, and an attribute for tracking the lineage of complex events. The event body specifies its *payload,* which we describe with a relational schema. For example, a purchase event would frequently contain the information of a purchase order ID. For our purposes payload is thought of merely as immediately available data, rather like a stack frame, and is opaque to the operator definitions. In other words, operator definitions are only concerned with the header information of events. Dot notation is used to refer to fields in event header (as well as payload). For example, Purchase.$V_s$ refers to the start valid time of the Purchase event. For an event type E, we use the notation e to denote a particular event instance of that type.

More specifically, we represent an event in the form (ID, $V_s$, $V_e$, $O_s$, $O_e$, $R_t$, cbt[]; p), where the first seven attributes represent the header information, and separated with the event body by a semi-colon, which payload, denoted as p, is specified. The first six attributes in the header are the same as the bitemporal schema. cbt[] is used to track the lineage of contributor events that form the composite event. The attribute cbt[] is a sequence (ordered set) of event references[2], and thus not in first order normal form. A sequence is denoted within square brackets. For example, we use [e1, e2,…,en] to denote that the value of cbt[] is a sequence of references to events e1 to en. In contrast, a set is specified within curly brackets. For example, {e1, e2,…,en} denotes a set of events e1 to en, where order is immaterial. For primitive events, the value of cbt[] is NULL.

---
[2] Event reference could be the pointer to that event or some other identifier.

### 3.3.2 Operators in WHEN Clause

We have introduced the notion of a canonical form R* for a bitemporal relation R previously. We now define a *shredded canonical form* as follows: Take R* as input. For each tuple e in R* with validity interval [$O_s$, $O_e$), replace it with $O_e$-$O_s$ tuples, such that all tuples have the same content as e in all attributes other than $O_s$ and $O_e$; their CEDR intervals are of length 1 but are all different; the union of these CEDR intervals is [$O_s$, $O_e$). We say e is shredded into these $O_e$-$O_s$ tuples. After shredding each tuple in R*, the resulting relation is a shredded canonical form. In defining the semantics of operators, we assume each input stream, a bitemporal relation, is in shredded canonical form. In all operator definitions, we require that the CEDR interval of all inputs is the same. This is a common condition we omit in the following definition of each operator.

In order to generate ID for the output events of an operator, we need a pairing function idgen, which takes a variable number of input IDs, and produces an ID. It has the property that the different sets of input IDs will generate different output IDs. In the output events, the value id for attribute ID is computed by idgen(e1.ID,…,ek.ID), where e1.ID through ek.ID are the set of input IDs. Also the value rt for attribute Rt in the output is the minimum root time value among all inputs e1 through ek. Note that how to assign $V_e$ value for outputs is in general orthogonal to the operator scope w. In the following operator definitions, we assume that $V_e$ of the output is set to e1.$V_s$+w, where e1 is the first contributor to the operator. Note that it is probably reasonable to set $V_e$ to infinity, or to the $V_e$ value of the last contributor of this operator.

**Event Sequencing –** The ability to synthesize events based upon the ordering of previous events is a basic and powerful event language construct. Almost all operators in the table below have a time based scope, denoted as w. A sequence based scope can be added if such functionality is required by any query CEDR wants to support.

| Operator | Description |
|---|---|
| **ATLEAST(n,E1,.,Ek, w)** | ATLEAST (n, E1, …, Ek, w) ≡ {(id, ein.$O_s$, ein.$O_e$, ein.$V_s$, ei1.$V_s$+w, [ei1, ei2, …, ein] ; ei1.p, ei2.p, …, ein.p) \| ei1.$V_s$<ei2.$V_s$<…<ein.$V_s$∧ein.$V_s$ – ei1.$V_s$ <= w∧{i1, i2, …, in} is a subset of {1, 2, …, k} ∧i1 != i2 != … != in}, where rt is the minimum root time value among ei1 through ein. |
| **ATMOST(n,E1,...,Ek, w)** | This operator is a syntactic sugar, which can be expressed with sliding window aggregate (count aggregate). In addition, it is possible to assign individual weights to contributors that can be used to adjust the counting. |
| ALL (E1, . . . , Ek, w) | ALL (E1, E2, . . . , Ek, w) ≡ ATLEAST (k, E1,E2,...,Ek, w) |
| ANY (E1,...,Ek) | ANY (E1, E2,...,Ek) ≡ ATLEAST (1, E1,E2,...,Ek, 1) |
| SEQUENCE(E1,,,Ek, w) | SEQUENCE(E1, E2, …, Ek, w) ≡ {id, $ek.O_s$, $ek.O_e$, $ek.V_s$, $e1.V_s+w$, rt, [e1, e2, …, ek]; e1.p, e2.p, …, ek.p) \| $e1.V_s<e2.V_s<…<ek.V_s \wedge ek.V_s - e1.V_s <= w$} |

Note that the correlation conditions in the definition of sequencing operators do not take root time into account. It can be easily made to do so if required by queries.

**Negation** – The event service can track the *non-occurrence* of an expected event, such as a customer not answering an email within a specified time. The negation feature has great utility in business processes.

Negation has to have a scope within which the non-occurrence of events is monitored. The scope can be time based or sequence based. For a time based scope, the start time of such a scope should be specified as well. For an efficient implementation, we first propose an operator UNLESS to implicitly specify such a start time, instead of allowing users to specify it. Informally, UNLESS(E1, E2, w) produces an output event when the occurrence of an E1 event is followed by no E2 event in the next w time units. The start time of the negation scope is therefore bound always to the occurrence (start valid time) of the E1 event. A variant UNLESS' that provides more flexible options for specifying the start time of the scope is then given. For sequence scope, we use operator NOT(E, SEQUENCE(E1, …, Ek, w)), where the second parameter of NOT, a sequence construct, is the scope for the non-occurrence of E. Since sequence scope is well specified within such a NOT operator, it is perfectly composable with other operators. For example, ALL(E1, NOT(E2, SEQUENCE(E3, E4, w')), w) produces an output when a sequence of E3, E4 events that occur within w' time units occurs within w time units of the occurrence of an E1 event, and between the E3 and E4 events there is no E2 event.

Finally, we propose the CANCEL-WHEN feature in CEDR, which is not found in existing systems. Event patterns normally do not "pend" indefinitely; conditions or constraints may be used to cancel the accumulation of state for a pattern (which would otherwise remain to aggregate with future events to generate a composite event). The CANCEL-WHEN construct is used to describe such constraints. CANCEL-WHEN (E1, E2) stops the detection for E1 when an E2 event occurs during the partial detection. Note the scope of E1 expressed by CANCEL-WHEN cannot in general be expressed by time or tuple based window in existing systems, since E2 could be a complex expression.

| Operator | Description |
|---|---|
| UNLESS(E1, E2, w) | UNLESS (E1, E2, w) ≡ {(e1.ID, $e1.O_s$, $e1.O_e$, $e1.V_s$, $e1.V_s+w$, e1.rt, [e1]; e1.p) \| there does not exist e2, such that $e1.V_s < e2.V_s < e1.V_s + w$} |
| UNLESS(E1,E2,n,w) | UNLESS' (E1, E2, w) ≡ {(e1.ID, $e1.O_s$, $e1.O_e$, $e1.V_s$, $e1.V_s+w$, e1.rt, max($e1.cbt[n].V_s+w$, $e1.V_s$), [e1]; e1.p) \| there does not exist e2, such that $e1.cbt[n].V_s < e2.V_s < e1.cbt[n].V_s + w$} <br><br> This operator allows users to specify that the start valid time of the negation scope for E2 is the n-th contributor to the E1 event. For this operator to be valid, at query compile time we need to check that the sequence specified by e1.cbt[] has length no less than n. Also, since the computation of E1 has its own scope, the $V_s$ field of the output of this UNLESS' operator should be set to the later one between the start valid time of E1 and the end of the negation scope for E2. <br><br> Whether we need such a flexible UNLESS' operator in CEDR is open to discussion. In the following discussion it is omitted. |
| NOT(E,SEQUENCE(E1,…,Ek,w)) | NOT(E,SEQUENCE (E1,…, Ek, w)) ≡ {es \| es is in SEQUENCE (E1,…, Ek, w) and there does not exist e, such that $es.cbt[1].V_s < e.V_s < es.cbt[k].V_s$} |
| CANCEL-WHEN (E1, E2) | CANCEL-WHEN (E1, E2) ≡ {e1 \| there does not exist e2, such that $e1.rt < e2.V_s < e1.V_s$} <br><br> Note that in this definition e2.rt is not involved. The definition can be changed to include this aspect. For example, $e1.rt < e2.rt < e2.V_s < e1.V_s$ is a reasonable definition as well. |

## 4. Consistency Guarantees

As stated earlier, due to unreliable (w.r.t. delivery order) network connections, stream events and their associated state changes may be delivered in non-deterministic order. In such situations, it can be highly undesirable to block until all the early data has provably arrived. Nevertheless, we can still produce output if we are willing to both retract incorrect output, and add the correct revised output. The ability to model and handle such *retractions* and insertions is a very important distinguishing feature of CEDR. This is modeled by moving to a tritemporal model, which adds a third notion of time, called CEDR time, denoted T. Figure 2 shows an example of a tritemporal **history table**.

**Figure 2. Example – Tritemporal history table**

| ID | $V_s$ | $V_e$ | $O_s$ | $O_e$ | $C_s$ | $C_e$ | ... | K |
|----|----|----|----|----|----|----|----|----|
| e0 | 1 | ∞ | 1 | 5 | 1 | 4 | | E0 |
| e0 | 1 | 10 | 5 | ∞ | 2 | 6 | | E1 |
| e0 | 1 | ∞ | 1 | 3 | 4 | ∞ | | E0 |
| e0 | 1 | 10 | 5 | 5 | 5 | ∞ | | E1 |
| e0 | 1 | 10 | 3 | ∞ | 6 | ∞ | | E2 |

Note that in this table, we still see the familiar valid time and occurrence time fields. In addition, we see a new set of fields associated with CEDR time. These new fields use the clock associated with an actual CEDR stream. In particular, $C_s$ corresponds to the CEDR server clock time upon event arrival. While critical for supporting retractions, CEDR time also reflects out of order delivery of data. Finally, note there is a K column, in which each unique value corresponds to an initial insert and all associated retractions, each of which reduce the $C_e$ compared to the previous matching entry in the table.

Figure 2 models both a retraction and a modification (described in Section 2) simultaneously, and may be interpreted as follows. At CEDR time 1, an event arrives whose valid time is [1,∞), and has occurrence time 1. At CEDR time 2, another event arrives which states that the first event's valid time changes at occurrence time 5 to [1,10). Unfortunately, the point in time where the valid time changed was incorrect. Instead, it should have changed at occurrence time 3. This is corrected by the following three events on the stream. The event at CEDR time 4 changes the occurrence end time for the first event from 5 to 3. Since retractions can only decrease $O_e$, the original E1 event must be completely removed so that a new event with a new $O_s$ time may be inserted. We therefore completely remove the old event from the system by setting $O_e$ to $O_s$. We then insert a new event, E2, with occurrence time [3, ∞) and valid time [1,10). Note that the net effect of all this is that at CEDR time 3, the stream, in terms of valid time and occurrence time, contains two events, an insert and a modification that changes the valid time at occurrence time 5. At CEDR time 7, the stream describes the same valid time change, except at occurrence time 3 instead of 5. Note, that retractions can be characterized and discussed using only occurrence time and CEDR time. Consequently, we will not discuss valid time or the ID fields further.

Before we proceed to defining our notions of consistency, we need to define a few terms. First, we define the notion of **canonical history table to time $t_o$** (occurrence time). This canonical form will be used later to describe a notion of stream equivalence. Two examples of non-canonical history tables are shown in Figure 3.

**Figure 3. Example – Two history tables**

| K | $O_s$ | $O_e$ | $C_s$ | $C_e$ | ... | K | $O_s$ | $O_e$ | $C_s$ | $C_e$ | ... |
|----|----|----|----|----|----|----|----|----|----|----|----|
| E0 | 1 | 5 | 1 | 3 | ... | E0 | 1 | ∞ | 1 | 2 | ... |
| **E0** | **1** | **3** | **3** | **∞** | ... | **E0** | **1** | **5** | **2** | **∞** | ... |

Putting a table into canonical form involves two steps. In the first step, called reduction, for each K, only the entry with earliest $O_e$ time is retained. The resulting history tables for the tables shown in Figure 3 are shown in Figure 4.

The next step, called truncation, changes any $O_e$ value in the table greater than $t_o$ to $t_o$. If there are any rows whose $O_s$ times are greater than $t_o$, they are removed. The canonical history tables for the tables shown in Figure 4, which were produced using truncation, are shown in Figure 5.

**Figure 4. Example – Two reduced history tables**

| K | $O_s$ | $O_e$ | $C_s$ | $C_e$ | ... | K | $O_s$ | $O_e$ | $C_s$ | $C_e$ | ... |
|----|----|----|----|----|----|----|----|----|----|----|----|
| **E0** | **1** | **3** | **3** | **∞** | ... | **E0** | **1** | **5** | **2** | **∞** | ... |

**Figure 5. Example – Two canonical history tables**

| K | $O_s$ | $O_e$ | $C_s$ | $C_e$ | ... | K | $O_s$ | $O_e$ | $C_s$ | $C_e$ | ... |
|----|----|----|----|----|----|----|----|----|----|----|----|
| **E0** | **1** | **3** | **3** | **∞** | ... | **E0** | **1** | **3** | **2** | **∞** | ... |

We define the notion of **canonical history table at $t_o$** (in place of "to $t_o$") as the canonical history table to $t_o$ with the rows whose occurrence time interval do not intersect $t_o$ removed. We are finally ready to define one of our most important terms, called logical equivalence:

**Definition 1**: Two streams $S_1$ and $S_2$ are **logically equivalent to $t_o$ (at $t_o$)** iff, for their associated canonical history tables to $t_o$ (at $t_o$), $CH_1$ and $CH_2$, $\pi_X(CH_1) = \pi_X(CH_2)$, where X includes all attributes other than $C_s$ and $C_e$.

Intuitively, this definition says that two streams are logically equivalent to $t_o$ (at $t_o$) if they describe the same logical state of the underlying database to $t_o$ (at $t_o$), regardless of the order in which those state updates arrive. For instance, the two streams associated with the two tables in Figure 3 are logically equivalent to 3 and at 3.

In order to describe our consistency levels, we have one more notion to define, a synchronization point. In order to define this, we need to describe an annotated form of the history table which introduces an extra column, called Sync. A table with such a column added is shown in Figure 6. The extra column (Sync) is computed as follows: For insertions Sync = $O_s$; for retractions Sync = $O_e$.

**Figure 6. Example - Annotated history table**

| K | Sync | $O_s$ | $O_e$ | $C_s$ | $C_e$ | ... |
|----|----|----|----|----|----|----|
| E0 | 1 | 1 | 10 | 0 | 7 | ... |
| **E0** | **5** | **1** | **5** | **7** | **10** | ... |

The intuition behind the Sync column is that it induces a global notion of out of order event arrival in CEDR. For instance: if and only if the global ordering of events achieved by sorting events according to $C_s$ is identical to the global ordering of events achieved by sorting events according to the compound key <Sync, $C_s$>, then there are no out of order events in the stream. Finally, we introduce the notion of a synchronization point, sync point for short:

**Definition 2**: A **sync point** w.r.t. an annotated history table AH is a pair of occurrence time and CEDR time ($t_o$, T), such that for each tuple e in AH, either $e.C_s <= T$ and e.Sync $<= t_o$, or $e.C_s > T$ and e.Sync $> t_o$.

Intuitively, a sync point is a point in both CEDR time and occurrence time which cleanly separates the past from the future in both time domains simultaneously. At these points in time, we have seen exactly the minimal set of state changes which can affect the bitemporal historic state up to occurrence time $t_o$. We now define our levels of consistency.

**Definition 3**: A standing query supports the **strong consistency** level iff: 1) for any two logically equivalent input streams $S_1$ and $S_2$, for sync points ($t_o$, $T_{S1}$), ($t_o$, $T_{S2}$) in the two output streams, the query output streams at these sync points are logically equivalent to $t_o$ at CEDR times $T_{S1}$ and $T_{S2}$. 2) for each entry E in the annotated output history table, there exists a sync point (E.Sync, $E.C_s$).

Intuitively, this definition says that a standing query supports strong consistency iff any two logically equivalent inputs produce exactly the same output state modifications, although there may be different delivery latency. Note that in order for a system to support this notion of consistency, the system must have "hints" that bound the effect of future state updates w.r.t. occurrence time. In addition, for n-ary operators, any combination of input streams can be substituted with logically equivalent streams in this definition. This is also true for the other consistency definitions and will not be discussed further.

**Definition 4**: A query supports the **middle consistency** level iff for any two logically equivalent input streams $S_1$ and $S_2$, for sync points ($t_o$, $T_{S1}$), ($t_o$, $T_{S2}$) in the two output streams, the query output streams at these sync points are logically equivalent to $t_o$ at CEDR times $T_{S1}$ and $T_{S2}$.

The definition of the middle level of consistency is almost the same as the high level. The only difference is that not every event is a sync point. Intuitively, this definition allows for the retraction of optimistic state at times in between sync points. Therefore, this notion of consistency allows us to produce early output in an optimistic manner.

**Definition 5**: A query supports the **weak consistency** level iff for any two logically equivalent input streams $S_1$ and $S_2$, for sync points ($t_o$, $T_{S1}$), ($t_o$, $T_{S2}$) in the two output streams, the query output streams at these sync points are logically equivalent at $t_o$ at CEDR times $T_{S1}$ and $T_{S2}$.

# 5. Consistency tradeoffs

In order to understand what these levels of consistency mean in a real system, we describe the role and functionality of a CEDR (logical) operator in a high level fashion.

**Figure 7. Anatomy of a CEDR operator**

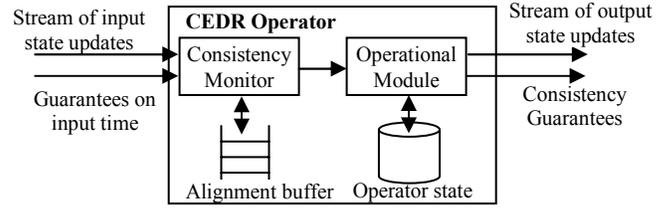

Similar to DSMSs, CEDR provides a set of composable operators that can be combined to form a pipelined query execution plan. Each CEDR operator, illustrated in Figure 7, has two components: a consistency monitor and an operational module. A consistency monitor decides whether to block the input stream in an alignment buffer until output may be produced which upholds the desired level of consistency. The operational module computes the output stream based on incoming tuples and current operator state.

Moreover, a CEDR operator accepts occurrence time guarantees on subsequent inputs (e.g. provider declared sync points on input streams). These guarantees are used to uphold the highest level of consistency, and allow us to reduce operator state in all levels of consistency. CEDR operators also annotate the output with a corresponding set of future output guarantees. These guarantees are fed to the next operator and streamed to the user with the query result.

An important property of CEDR operators is that we use formal descriptions of operator semantics to prove that at common sync points, operators output the same bitemporal state regardless of consistency level. As a result, one can seamlessly switch from one consistency level to another at these points, producing the same subsequent stream as if CEDR had been running at that consistency level all along.

**Figure 8. Consistency tradeoffs**

| Consistency | Orderliness | Blocking | State Size | Output Size |
|---|---|---|---|---|
| Strong | High | Low | Low | Minimal |
| | Low | High | High | Minimal |
| Middle | High | None | Low | Low |
| | Low | None | High | High |
| Weak | High | None | Low- | Low- |
| | Low | None | Low- | Low- |

Figure 8 illustrates the qualitative implications of running CEDR at a specific consistency level. The table considers two cases per consistency level: a highly-ordered stream and a very out-of-order stream, where orderliness is measured in terms of the frequency of application declared *sync point*.

Figure 8 shows that the middle and strong consistency levels have the same state size – the tradeoff here is between blocking times (responsiveness) and the output size. This is caused by the contrasting way that the two levels handle out of order events. The strong level aligns tuples by blocking, possibly resulting in significant blocking and large state, if the input is significantly out of order. In contrast, the middle level optimistically generates output, which can be repaired later using retractions and insertions. Since these retractions can affect output as far back in time as the last sync point, the middle level must maintain the same state as the strong level to generate the necessary retractions in all cases.

Both the middle and the weak consistency levels are non-blocking – they are distinguished by their output correctness up to (versus at) arbitrary points of time. More specifically, in the weak consistency level, we are not always obligated to fix earlier state, and may therefore "forget" some events which arrived since the last sync point. As a result, when events are highly out of order, both output size and state size are much improved over the middle level. When events are ordered, the strong level of consistency may be enforced with marginal added cost over weak and middle consistency.

It is worth noting, the ability to both remember and block do not have to be *all or nothing* properties of our operators. Rather, one can limit blocking and memory to specific lengths of either application or CEDR time. This leads to the infinite spectrum of consistency levels described in Figure 9, which shows the space of valid consistency levels where the maximum memory time M is one dimension, and maximum blocking time B is the other dimension.

**Figure 9. Consistency tradeoffs**

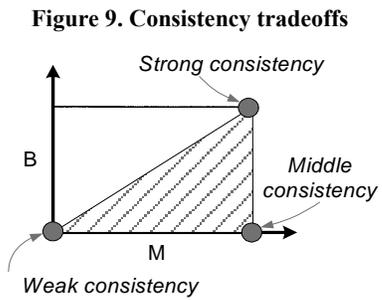

The interesting part of the spectrum is the lower right triangle since increasing the maximum blocking time beyond the maximum memory time has no effect on operator behavior. Note that the lower left corner of the triangle corresponds to the weakest possible consistency level, which is both non-blocking and memoryless. As we travel along the X-axis of the graph, we are willing to remember progressively further and further into the past, but remain non-blocking. At the extreme, we are willing to remember everything, and are therefore at the middle level of consistency at the lower right (at infinity) corner of the triangle. From this corner, we proceed up to the top right corner, where we are willing to both block arbitrarily long and remember everything if need be. This obviously corresponds to the highest possible level of consistency.

## 6. Run-time Operator Semantics

In CEDR, run-time operator semantics are "pure" in the sense that the result of a CEDR standing query must be ultimately unaffected by temporary stream states that are caused by out of order event arrival as well as retractions. More formally, a properly specified CEDR operator must be well behaved according to the following definition:

**Definition 6**: A CEDR operator O is **well behaved** iff for all (combinations of) inputs to O which are logically equivalent to infinity, O's outputs are also logically equivalent to infinity.

Intuitively, the above definition says that a CEDR operator is well behaved as long as the output produced by the operator semantically converges to the output produced by a perfect version of the input without retractions and out of order delivery.

Also, since the above definition induces input stream equivalence classes based on logical equivalence, we need only to define operator semantics on the infinite canonical history tables with the CEDR time fields projected out. We will call these tables **ideal history tables**, By defining operators using ideal history tables, we ensure that for each equivalence class, we define operator semantics on the equivalence class member which excludes retractions and out of order delivery. It is up to the implementations of individual operators, which is beyond the scope of this paper, to uphold logically equivalent operator output behavior for all logically equivalent inputs.

While a fully realized set of CEDR operators would support both retractions and modifications, the discussion in this section would be less relevant to existing systems if we defined our operators in this manner. We will therefore, in this section, assume that there are no modifications, and that the occurrence and valid time fields are merged into one valid time field, whose lifetime may be shortened using retractions. All the reasoning and definitions in Section 4 are, in this context, in terms of valid time and CEDR time instead of occurrence time and CEDR time. Furthermore, in this context, the resulting ideal history tables have only one temporal dimension (valid time) and are therefore called **unitemporal ideal history tables**. We leave it as a technical challenge to define precisely the semantics of our operators in the presence of modifications.

Summing up, the semantics of our operators are defined on the unitemporal ideal history tables of the inputs, such as the one shown in Figure 10. In all definitions, we refer to the input streams as $S_1,…,S_m$, and the set of events in each associated unitemporal ideal history table as $E(S_i)$. Each individual event has the fields shown in Figure 10.

**Figure 10. Example – Unitemporal ideal history table**

| ID | $V_s$ | $V_e$ | Payload |
|---|---|---|---|
| E0 | 1 | 5 | P1 |
| E1 | 4 | 9 | P2 |

The output of the operator is described as the set of events in the unitemporal ideal history table of the output. Each element of the output is therefore described as the triple ($V_s$, $V_e$, Payload). We begin with the definitions of operators which will be very familiar to the readers of this paper:

**SQL projection** is a generalization of the relational projection operator, in that we can specify an arbitrary function f to transform the payload of each input tuple. Consequently, the output payload schema may be different from the input payload schema. Note that f cannot affect the timestamp attributes. SQL projection is defined as follows:

**Definition 7**: SQL projection $\pi_f(S)$:
$\pi_f(S) = \{(e.V_s, e.V_e, f(e.Payload)) \mid e \in E(S)\}$

**Selection** corresponds exactly to relational selection. It takes a boolean function f which operates over the payload. The definition follows:

**Definition 8**: Selection $\sigma_f(S)$:
$\sigma_f(S) = \{(e.V_s, e.V_e, e.Payload) \mid e \in E(S) \text{ where } f(e.Payload)\}$

Similarly, the next operator is **join**, which takes a boolean function f over two input payloads:

**Definition 9**: Join $\bowtie_{f(P1,P2)}(S_1, S_2)$:
$\bowtie_{\theta(P1,P2)}(S_1, S_2) = \{(V_s, V_e, (e_1.Payload \text{ concatenated with } e_2.Payload)) \mid e_1 \in E(S_1), e_2 \in E(S_2), V_s = \max\{e_1.V_s, e_2.V_s\}, V_e = \min\{e_1.V_e, e_2.V_e\}, \text{ where } V_s < V_e, \text{ and } \theta(e_1.Payload, e_2.Payload)\}$

Intuitively, the definition of join semantically treats the input streams as changing relations, where the valid time intervals are the intervals during which the payloads are in their respective relations. The output of the join describes the changing state of a view which joins the two input relations. In this sense, many of our operators follow view update semantics such as those specified in [10].

We include in our algebra a number of other operators, such as **union, difference**, **groupby,** and aggregates such as **max**, **min**, and **avg**. These operators all follow view update semantics, and since their relational counterparts are well understood we do not give formal definitions here. Instead, we discuss an attribute which all operators discussed so far have in common, called **view update compliance**.

Before we can define view update compliance, we need to first introduce some other terminology:

**Definition 10**: meets ($I_1$, $I_2$), coalesce ($E_1$, $E_2$), *(S):
Two intervals $I_1=[T_1, T_2]$, $I_2=[T_1', T_2']$ **meet** iff $T_2 = T_1'$

Two events can be **coalesced** if their payloads are the same and their associated valid time intervals meet. Two coalesced events $e_1=(V_s, V_e, P)$, $e_2=(V_s', V_e', P)$ are replaced with a single event $e=(V_s, V_e', P)$.

The * operator on a stream returns the unitemporal history table that results from the repeated application of coalescence to the unitemporal ideal history table until coalesce cannot be applied further:

We are now ready to define relational view compliance:

**Definition 11**: A unary CEDR operator O is view update compliant iff for all R, S s.t. *(R) and *(S) are identical, *(O(R)) and *(O(S)) are also identical

Intuitively, the above definition states that semantically, an operator must be insensitive to the way that changes in state are packaged. This is why, for instance, the operator must treat a payload whose lifetime is chopped into several insert events the same way as a payload whose lifetime is described in one event with a larger, equivalent lifetime.

The above definition may be generalized in the obvious way to n-ary operators. In addition, this definition assumes that the underlying streams model relations, and therefore don't allow duplicate payloads with overlapping valid time intervals. A more general definition could be crafted to handle bag semantics for the underlying relations.

Unsurprisingly, most streaming systems (e.g. [5], [10]) implement operators that are view update compliant. What is interesting is that the features which are considered unique to streams, like windows, and the separation of inserts and deletes, are not view update compliant, which raises the question: What non-view update compliant operators are necessary in a streaming system? What guarantees should they uphold?

We will therefore introduce our one non-view update compliant operator, **AdjustLifetime**, using this simple, but powerful operator we can build many windowing constructs and separate inserts from deletes. It is worth noting that AdjustLifetime, while non-view update compliant, is well behaved. AlterLifetime takes two input functions $fV_s(e)$ and $f\Delta(e)$. Intuitively, Alterlifetime maps the events from one valid time domain to another. In the new domain, the new $V_s$ times are computed from $fV_s$, and the durations of the event lifetimes are computed from $f\Delta$. One could therefore regard this operator as a constrained form of project on the temporal fields. The precise definition follows:

**Definition 12**: AlterLifetime $\Pi_{fvs, f\Delta}(S)$
$\Pi_{fvs, f\Delta}(S) = \{(|f_{Vs}(e)|, |f_{Vs}(e)| + |f_\Delta(e)|, e.Payload) \mid e \in E(S)\}$

We now define a moving window operator, denoted W, as a special instance of Π. This operator takes a window length parameter wl, and clips the validity interval of its input based on wl. More precisely: $W_{wl}(S)=\Pi_{Vs, min(Ve-Vs, wl)}(S)$

One can similarly define hopping windows using integer division. Finally, we can even use the AlterLifetime operator to easily get all inserts and deletes from a stream:

Inserts(S)= $\Pi_{Vs, \infty}(S)$

Deletes(S)= $\Pi_{Ve, \infty}(S)$

## 7. Conclusions and Future work

In this paper, we have presented a number of challenges for existing streaming systems. These challenges include the ability to handle negation, event selection and consumption, application driven modifications, and out of order event delivery in a principled, flexible manner. In order to address these challenges we propose a powerful temporal stream model. We build on this model in a number of ways:

1. We formally define the notion of retractions, which can be used to describe a spectrum of possible system behaviors and performance tradeoffs in response to out of order delivery of data.

2. We provide denotational semantics for a set of streaming run-time operators, most are view update compliant, and all are well behaved. The only operator which is not view update compliant is a simple, but novel streaming operator which can be used to implement a plethora of window types and the separation of inserts and deletes.

Ongoing work in the CEDR project is proceeding in a number of research directions. One effort is to complete the set of compilation rules from our language to our run-time operator algebra. In addition, we are working on algorithms which efficiently implement our algebra across the full spectrum of consistency levels. Another interesting direction is optimization and query rewrite rules. For instance, we are considering consistency sensitive query optimizations that when permissible, can determine when to switch from one consistency level to another under periods of heavy load due to event bursts.